\documentclass[draft]{article}
\usepackage{amsmath,amsfonts,latexsym, amssymb}

\usepackage{color}

\newcommand{\Tr}{\mbox{{\rm Tr}\,}}

\newcommand{\e}{\varepsilon}

\newcommand{\rd}{{\rm d}}
\newcommand{\bR}{{\mathbb R}}
\newcommand{\bbZ}{{\mathbb Z}}

\newcommand{\bs}{ \begin{split} }
\newcommand{\es}{ \end{split} }

\newcommand{\bN}{{\mathbb N}}

\newcommand{\be}{\begin{equation}}
\newcommand{\ee}{\end{equation}}

\newcommand{\om}{{\omega}}

\newcommand{\C}{{\mathbb C}}
\newcommand{\Z}{{\mathbb Z}}

\newcommand{\R}{{\mathbb R}}
\newcommand{\N}{{\mathbb N}}

\newcommand{\wt}{\widetilde}

\newtheorem{theorem}{Theorem}
\newtheorem{corollary}[theorem]{Corollary}
\newtheorem{lemma}[theorem]{Lemma}
\newtheorem{proposition}[theorem]{Proposition}

\newtheorem{remark}[theorem]{Remark}
\newtheorem{definition}[theorem]{Definition}
\newcommand{\qed}{\hfill\fbox{}\par\vspace{0.3mm}}
\newenvironment{proof}{{\bf Proof.}} {\hfill\qed}

\numberwithin{equation}{section}
\numberwithin{theorem}{section}

\title{Anderson Localization at Band Edges for Random Magnetic
Fields}
\author{L\'aszl\'o Erd\H os\thanks{Partially supported by SFB-TR12 of
the German Science Foundation},\; David Hasler
 \\
\\
Institute of Mathematics, University of Munich, \\
Theresienstr. 39, D-80333 Munich, Germany \\
\text{lerdos@math.lmu.de, hasler@math.lmu.de} }

\date{March 18, 2011}

\begin{document}

\maketitle

\begin{abstract} We consider a magnetic Schr\"odinger operator in two dimensions.
The magnetic field is given as the sum of a  large and constant magnetic field and
 a  random magnetic field.
Moreover, we allow for an  additional  deterministic potential as well as  a magnetic
 field which are both periodic.
We show that the spectrum of this operator is contained in
broadened bands around the Landau levels and  that the edges of these bands consist of pure point spectrum with
exponentially decaying eigenfunctions. The proof is based on a recent Wegner estimate obtained in \cite{EH2} and a multiscale analysis.
\end{abstract}


\medskip

{\it Running title:} Magnetic localization at band edges

\medskip


\section{Introduction}

The energy levels of a spinless
 quantum particle in the two dimensional
Euclidean space $\bR^2$ subject to a constant
magnetic field $B_0$ are given by the Landau
levels, $(2n+1)B_0$, $n=0,1,2,\ldots$.
A perturbation with an inhomogeneous random
stationary magnetic field broadens the Landau
levels into spectral bands.
In this paper we prove Anderson localization
near the band edges and we thus generalize
our previous work \cite{EH2} that treated only
the bottom of the spectrum.

In the standard model for Anderson localization with a magnetic field
(see, e.g., \cite{CH, DMP, FLM, GK, W})
the random perturbation is given by an additive external potential.
In our model  the randomness is carried by the magnetic field.
The main mathematical difference between these models is twofold.

First, the correlation structure of the  local Hamiltonians
for magnetic fields is much more involved.
Assuming finite range correlations for the random perturbation, in case of
potential perturbations the local Hamiltonians
on distant domains are  independent since
the external potential acts locally.
Some sufficiently decaying but not finite range correlations can
also be treated with the known methods, see  \cite{DK,kirstosto98}
and references therein.
In case of magnetic perturbations, it is the vector potential $A$
and not the magnetic field $B=\nabla \times A$
that appears directly in the Hamiltonian.
Since the dependence of the vector potential on the
magnetic field is nonlocal, the local Hamiltonians
with distant domains are typically strongly correlated
even for magnetic fields with a short range correlation.
This strong correlation cannot be directly
tackled with the standard methods of multiscale analysis,
but using appropriate gauge transformations helps.

The second  difference between
random external potentials and random magnetic fields
is that the energy depends  monotonically on
the external potential but not on the magnetic field.
A cornerstone of any existing proof of Anderson
localization is the Wegner estimate whose standard proofs rely
on monotonicity. Prior to our work \cite{EH2}, Wegner
estimate, and hence localization,
 has only been proven for random magnetic fields
with a zero flux condition \cite{KNNN} and
for fields generated by stationary vector potentials
in \cite{U,GHK}, motivated by a method in \cite{HK}.
Note that stationary vector potentials imply that the flux is zero on average.

In  \cite{EH2} we developed a new method to
prove  Wegner estimate for stationary random fields, i.e.
for a model without monotonicity. In particular, zero-flux condition
was not needed.  We also proved Lifshitz
tail at the bottom of the spectrum. These ingredients,
combined with the usual multi-scale argument yielded localization
at the bottom of the spectrum. In \cite{EH3}  we solved the same problem for
the lattice model.

In the current paper we extend our method for higher band edges.
The Wegner estimate and the multi-scale argument remain
essentially unchanged and we will just quote the necessary results.
The new ingredients are (i) the precise location of the
higher band edges and (ii) an estimate on the Lifshitz tail.
Both results are especially effective if the
background constant field is strong compared with
the random perturbation.

\section{Model and Statement of Results}\label{sec:field}
\label{sec:thm}

We work in $\bR^2$ and we set $|x|_\infty := \max\{ |x_1|, |x_2|\}$
for any $x\in \bR^2$. We shall denote magnetic fields by $B$.
Let $A$ be a magnetic vector potential such that $\nabla \times A = B$.
By $H(A)$  we
denote the magnetic Schr\"odinger operator on $L^2(\R^2)$  with a
bounded external potential $V$, i.e.,
$$H(A) = (p - A)^2  + V .$$
We realize this as a self adjoint operator by means of the Friedrichs extension.
If we refer to statements which are independent of the particular choice of gauge,
with a slight abuse of notation, we shall
occasionally write $H(B)$. In particular,
we denote by $\sigma(H(A))$ the spectrum of the magnetic Schr\"odinger operator $H(A)$.
Since the spectrum is gauge invariant
sometimes we will also use the notation $\sigma(H(B))$.

We consider a deterministic magnetic field
$B_{\rm det}(x) = B_0 + B_{\rm var}(x)$ where $B_0$ denotes  a constant magnetic field
and $B_{\rm var}$ is a perturbation that typically varies in space.
We perturb this deterministic magnetic field by a random one, i.e., we consider
\be
B =     B_\omega= B_{\rm det}  + \mu B_{\rm ran}^\omega,
\label{B}
\ee
where $\mu \in (0,1]$ denotes the coupling constant and $B_{\rm ran}^\omega$
is a random magnetic field constructed as follows.

We choose a  profile function $u\in C_0^1(\bR^2)$, $0\le u\leq 1$.
Fix $k\in \bN$  and define the lattice $\Lambda^{(k)} = (2^{-k}\bbZ)^2$.
For $z\in \Lambda^{(k)}$ define
\be
       \beta_z^{(k)}(x) :=  u\big(2^{k}(x-z)\big) .
\label{betadef}
\ee
The randomness is represented by a  collection of
independent random variables
$$
\omega = \{ \om^{(k)}_z\; : \; k\in\bN,
z\in \Lambda^{(k)}\} .
$$
We assume that all $\omega^{(k)}_z$ have zero
expectation, and they satisfy a bound that is uniform in $z$
\be
     | \om^{(k)}_z| \leq \sigma^{(k)}:=  C_{\rm ran} e^{-\rho k} ,
\label{lowbound1}
\ee
with some $\rho>0$.
By  $v^{(k)}_z$ we shall denote the density function
of $\om^{(k)}_z$ (which strictly speaking might be a distribution).
For each  $(k,z)  \in \mathcal{L} :=
\bigcup_{k \in \N} \{ k \} \times \Lambda^{(k)} $ we have a
 probability measure with
density $v_z^{(k)}$. The associated product measure, ${\mathbb P}$,
 is probability measure on
$\Omega = \R^{\mathcal{L}}$, and we denote expectation
with respect to this measure
by ${\mathbb E}$.
We define the random magnetic field as
\be
 B_{\rm ran}^\omega(x) =  B_{\rm ran}(x) := \sum_{k=0}^\infty  B^{(k)}(x), \qquad
B^{(k)}(x) := \sum_{z\in \Lambda^{(k)}} B^{(k)}_z(x), \qquad
     B^{(k)}_z(x) := \om^{(k)}_z   \beta_z^{(k)}(x) \; ,
\label{bconst}
\ee
i.e. $B_{\rm ran}^\omega$ is the sum of independent local magnetic fields on each scale
$k$ and at every $z \in \Lambda^{(k)}$.

The random magnetic field just constructed  will in general live
on infinitely many scales. This structure will be necessary
to prove the Wegner estimate and hence the Anderson localization.
Before that, we will state several results about the
location of the spectrum. In these results the genuine multi-scale
structure is not necessary;  it is allowed that
there is only one scale. This case is included in the
above construction by choosing
 all $\om^{(k)}_z = 0$ for $k \ge 1$
(which corresponds to the case where the corresponding
distribution  is  a point measure at the origin).

To state our results about the deterministic spectrum we need  that the random
magnetic field   is stationary. This is ensured if  we  make  the following
assumption.

\begin{itemize}
\item[\bf (i.i.d.)]   For any fixed $k \in \N$ the random variables
 $\{ \om^{(k)}_z : z   \in \Lambda^{(k)}  \}$
are  identically distributed.
\end{itemize}

\begin{theorem} \cite[Theorem 3.2]{EH2} \label{thm:deterministic} Suppose $B_\omega$ is a random magnetic field a
random magnetic field  constructed   in \eqref{B}, \eqref{betadef}, and \eqref{bconst},
satisfying \eqref{lowbound1} and {\bf (i.i.d.)}. Assume $B_{\rm var}$ and $V$ are $\Z^2$-periodic.
Then there exists a
set $\Sigma \subset \R$  and  a set
 $\Omega_1 \subset \Omega$ with  ${\mathbb{P}}(\Omega_1) = 1$ such that for all $\omega \in \Omega_1$
$$
\sigma( H(B_\omega)) = \Sigma.
$$
\end{theorem}

Henceforth we will denote by  $\Sigma$  the almost sure  deterministic spectrum of  $H(B_\omega)$.
The next two theorems provide estimates on the location of  the deterministic spectrum.
We define  two specific configurations of the collection of random variables,
\be
 ( \omega_+)_z^{(k)} := {\rm ess \,  sup} [ \omega_z^{(k)}]_+ , \quad ( \omega_-)_z^{(k)} :=  - {\rm ess \, sup} [ - \omega_z^{(k)}]_+ ,
\ee
where $[f]_+ = \max(0,f)$ denotes the positive part. The configuration
$\omega_{\pm}$ corresponds to maximal respectively minimal magnetic field.
These configurations give rise to the following points close to the band edges of the
deterministic  spectrum (provided one has band structure),
\be\label{Esupold}
E_{n}^{-} :=  \inf_{x \in \R^2} \left[  (2 n + 1 ) B_{\omega_-}(x) + V(x) \right] , \quad
E_{n}^{+} := \sup_{x \in \R^2} \left[ ( 2 n + 1 ) B_{\omega_+} (x) + V(x) \right].
\ee
Moreover,  we need that the derivatives of the fluctuations are not to large. This is quantified in
terms of the following constants
\begin{align}
&K_2^{\pm} :=  \| \nabla B_{\omega_{\pm}}\|_\infty  + \|\nabla V \|_\infty ,  \label{eq:k2pmdef} \\
& K_3^{\pm}  = \| \nabla B_{\omega_{\pm}} \|_\infty^2 . \label{eq:k3pmdef}
\end{align}
Note that using \eqref{lowbound1} and the support properties of the profile function $u$
  for any $\omega$ in the support of ${\mathbb P}$ we have
\be \label{eq:boundonB}
 \| \nabla B^\omega_{\rm ran}\|_\infty \le
 C \sum_k e^{-\rho k} 2^k \|\nabla u\|_\infty \le (1- 2e^{-\rho})^{-1}
 \|\nabla u\|_\infty
\ee
i.e.
 a sufficient condition for $B^\omega_{\rm ran}$ to be  differentiable is $\rho > \ln 2$.

\medskip

We will use  the following theorem in the proof of our localization
result to show that the interval for which we prove pure
point spectrum, does in fact contain spectrum, \eqref{eq:thereisspec} in Theorem \ref{thm:loclization}.

\begin{theorem} \label{thm:estonspeca}  Suppose $B_\omega$ is a random magnetic field
constructed   in \eqref{B}, \eqref{betadef}, and \eqref{bconst},
satisfying \eqref{lowbound1}, {\bf (i.i.d.)}, and  $0 < b_0 \leq  B_{\omega_{\pm}}$
with some positive constant $b_0$.
 Assume $B_{\rm var}$ and $V$ are $\Z^2$-periodic.
Then there exists  a  $C_{{\rm int},n}$  such that
\begin{align} \label{eq:estonbandedge}
{\rm dist}( \Sigma , E_{n}^{\pm}) \leq  C_{{\rm int}, n} 
[ K_2^\pm b_0^{-1/2}  + K_3^\pm b_0^{-2} ]  .
\end{align}
\end{theorem}

The following theorem can be used to establish  that the random Hamiltonian
exhibits band structure  under the additional condition that $B_{\rm det}= B_0$
is constant, i.e. $B_{\rm var}=0$ and $V=0$.

\begin{theorem} \label{thm:estonspecb}  Let  $B_{\rm det}= B_0$
be constant and $V=0$.
Suppose $B_\omega$ is a random magnetic field
 constructed   in \eqref{B}, \eqref{betadef}, and \eqref{bconst},
satisfying \eqref{lowbound1}, {\bf (i.i.d.)},
and $0< b_0 \leq B_\omega  \leq K_0 b_0$ for some   $K_0>1$. Assume that the essential
support of $v_z^{(k)}$ is  an interval (possibly consisting of a single point).
Then there exist intervals
$\Sigma_n =[\Sigma_n^-,\Sigma_n^+]$ containing $B_0 (2 n + 1)$ such that
$$
\Sigma =  \bigcup_{n=0}^\infty \Sigma_n .
$$
There exists a constant $C_n$ (depending only on $n$ and $K_0$) such that
$$ {\rm dist}( \Sigma_n^\pm , E_{n}^{\pm})  \leq  C_n  [ K_2^\pm b_0^{-1/2}  + K_3^\pm  b_0^{-2} ]  . $$
\end{theorem}
\medskip

\begin{remark}\label{rmk:band}
 Theorem \ref{thm:estonspecb} can be used to show that the spectrum exhibits band structure.
For fixed $n$  and bounded random field $| B_{\rm ran}| \leq C$ the intervals $\Sigma_n$
will be disjoint for large $b_0$.
However,  the distance between  neighboring Landau levels does not depend on $n$, but
the width of $\Sigma_n$ is  typically increasing in  $n$. Thus
for fixed $b_0$ and  large $n$  the intervals $\Sigma_n$ will no longer  be disjoint and
the gaps close, i.e. the high-energy spectrum will be a half-life.
\end{remark}

\medskip

Below we list additional assumptions which are needed in \cite{EH2} to show the
Wegner estimate, and which are hence needed to prove localization.
It is here where we need that the random magnetic field lives on
arbitrarily small scales.
The  profile function  satisfies one of the following two conditions for some
sufficiently small $\delta$:
either
\be
   u(x)\equiv 0  \quad \mbox{for}
   \quad |x|_\infty  \geq \frac{1}{2}+\delta
    \quad\mbox{and}\quad     u(x) \equiv 1, \quad \mbox{for}
   \quad |x|_\infty  \leq
   \frac{1}{2} - \delta
\label{kcond}
\ee
or
\be
    u(x) = \delta^2u_0( x\delta)
  \;\;\; \mbox{with some $u_0\in C^1_0(\bR^2)$},\;\; \int_{\bR^2} u_0=1,
\quad
u_0(x)\equiv0\;\;
\mbox{for $|x|_\infty \ge 1$}.
\label{kcond1}
\ee
In both cases $\delta$ can be chosen as a sufficiently small
positive number $\delta\leq\delta_0\leq 1$, and  the threshold $\delta_0$
can be chosen as
\be
\begin{split}
  \delta_0 =&\frac{1}{3200} \quad \mbox{under condition \eqref{kcond}}
\label{delta0}
\cr
\delta_0 = & \frac{1}{640+32\|\nabla u_0\|_\infty^2}
    \quad \mbox{under condition \eqref{kcond1}}.
\end{split}
\ee
The density function  $v_z^{(k)}$ is in $C^2_0(\bR)$
and  satisfies
\be
    \int_{\bR} \Big| \frac{\rd^2 v^{(k)}_z}{\rd s^2}(s)\Big|
  \rd s \leq C[\sigma^{(k)}]^{-2} = Ce^{2\rho k},
\label{2der}
\ee
 in particular the magnetic field has a non-trivial random component on every scale.
The profile function satisfies
\begin{equation} \label{eq:profileidentity1}
  U(x): =  \sum_{z \in \Z^2} u(x - z) \geq c_u  \qquad \mbox{and}
 \quad  {\rm sup}_{x \in \R^2} U(x)   = 1 ,
\end{equation}
for some  positive constant $c_u >0$.
We are given positive numbers $b_0$  and $K_0 > 3$, such that
\be
0 < 2 b_0\leq  B_{\rm det}(x) \leq (K_0 - 1 ) b_0 ,
\label{detbound}
\ee
Moreover, we assume that
\be \label{newcondranb}
    \sum_{k=0}^\infty \sigma^{(k)}\leq b_0,
\ee
i.e.$(1-e^{-\rho})b_0 \geq C_{\rm ran}$. It follows that
\be
    0< b_0 \leq B_\omega  \leq K_0 b_0.
\label{lowbound}
\ee
 We also assume the following condition on the potential
\be \label{eq:conV}
\| V  \|_\infty \leq b_0/4 .
\ee

The following  list summarizes   the assumptions  
for which a Wegner estimate 
was shown in \cite{EH2} (see Theorem~\ref{thm:wegner} below for the precise statement).

\medskip
\begin{itemize}
\item[\bf ($\boldsymbol{\rm W}$)] $B_\omega$ is a random
magnetic field  constructed   in \eqref{B}, \eqref{betadef}, and \eqref{bconst},
satisfying \eqref{lowbound1}, \eqref{kcond} or \eqref{kcond1} for some  $0 < \delta \leq \delta_0 $,
and \eqref{2der}--\eqref{newcondranb} hold with some parameters $K_0>3$, $b_0>0$
and  $\rho >  \ln 2$. Let also \eqref{eq:conV} hold.
\end{itemize}
\medskip

Finally, we now collect the additional assumptions needed to prove the result about localization.
Let $\tau>0$ be a parameter.

\medskip
\begin{itemize}
\item[\bf ($\boldsymbol{{\rm A}_\tau}$)]
  Hypothesis  {\bf ($\boldsymbol{\rm W}$)}  and {\bf (i.i.d.)} hold,
 and  $B_{\rm var}$ and $V$ are $\Z^2$-periodic.
 The density function $v^{(0)} = v^{(0)}_z$  is supported on an interval and
 there exists a polynomial bound on its lower tail, in the sense that
  there exist constants $c_v > 0$
 such that for  all $h \geq 0$  the probability distribution functions
\be \label{eq:probdist1}
 \nu_+(h) := \mathbb{P}( \omega^{(0)} \leq  \omega_+^{(0)} - h ) , \quad
  \nu_-(h) := \mathbb{P}( \omega^{(0)} \geq  \omega_-^{(0)} + h )
\ee
satisfy
\be
 \nu_{\pm}(h) \leq c_v h^\tau .
\ee
\end{itemize}

\medskip

\noindent
To state the result we introduce the following quantity
\begin{align*}
 K_2 := {\rm ess \, sup}_\omega \| \nabla B_\omega\|_\infty + 
\| \nabla V \|_\infty .
\end{align*}
Observe that one can derive an explicit upper bound on
 this constant with the
help of  \eqref{eq:boundonB}.

\medskip

\begin{theorem} \label{thm:loclization}  Let
 {\bf ($\boldsymbol{{\rm A}_\tau}$)} hold for some  $\tau > 2 $
and let $K_0$, $b_0$ be the parameters in  ($\boldsymbol{\rm W}$).
Let $B_\omega = B_{\rm det}+ \mu B_{\rm ran}^\omega$ with $\mu\in (0,1]$ be the random
magnetic field with a vector potential  $A_\omega$.
 For every $n \in \N$ there exists an $\e_n>0$ and 
$U_n > 0$ (independent of $b_0$
but depending on $K_0$, $\tau$, $\rho$, $\delta$, $C_{\rm ran}$, $c_u$, $c_v$)
 such that for any $b_0$ with  $ b_0\ge U_n (K_2^4+1)$  the interval
\be \label{eq:intofloc}
I_n: = [E_{n}^{+} - \e_n , E_{n+1}^{-}+ \e_n ]
\ee
is non-emtpy and 
 for almost every $\omega$ the operator  $H(A_\omega)$ has in $I_n$
pure point spectrum with exponentially decaying eigenfunctions.
 Moreover,  this interval  contains spectrum
at least near its edges, i.e.
\be \label{eq:thereisspec}
(E_{n}^{+} - \e_n , E_{n}^{+} + \e_n  )   \cap \Sigma \neq \emptyset , \quad
 ( E_{n+1}^{-} - \e_n  , E_{n+1}^{-} + \e_n )  \cap \Sigma \neq \emptyset.
\ee
\end{theorem}

\begin{remark} 
If  $B_{\rm var} = V = 0$, then it is a trivial consequence of
 Theorem \ref{thm:estonspecb}  that both intersections  \eqref{eq:thereisspec} contain
in fact an interval of nonzero length.
\end{remark}

\begin{remark}
For a typical random magnetic field we have  $E_{n+1}^{-} <  E_{n}^{+}$
 for any sufficiently
large $n$  (the lower threshold  depends on $b_0$).
 This is the signature that the broadening
of the Landau levels will overlap, see Remark~\ref{rmk:band}.
\end{remark}

\begin{remark} 
We note that the  assumption   {\bf (i.i.d.)} could be relaxed to the  weaker assumption that  for any fixed $k \in \N$ and all
$z \in \Lambda^{(k)}$, $\{ \omega_z^{(k)} : z' = z + w , w \in \Z^2 \}$ are identically distributed.
 Our proofs
show that the results of  Theorems  \ref{thm:estonspecb},   \ref{thm:estonspecb},  and \ref{thm:loclization} still hold under this weaker assumption.
\end{remark}

\medskip

We will use the notation that ${\bf 1}_S$ as well as   $\chi_S$  denotes the
characteristic function of a set $S$.

\medskip

Let us now outline the remaining part of the paper.
In Section \ref{sec:outerbound}, we prove a result, stated in  Theorem \ref{thm:boundonspec}, which gives an outer bound on the spectrum
of a magnetic Hamiltonian. It will be used  in the  proof of  both
 Theorem \ref{thm:estonspecb}
and Theorem \ref{thm:loclization}. The result might be of interest of its own.

In Section \ref{sec:detspec}, we prove Theorems \ref{thm:estonspeca} and \ref{thm:estonspecb}.  Theorem \ref{thm:estonspeca} will follow by choosing a
suitable  trial state. To prove
 Theorem \ref{thm:estonspecb} we will use the outer bound on the  spectrum  and a perturbation theory
 argument  in the  continuous spectrum.

Section \ref{sec:loc} is devoted to the proof of  Theorem \ref{thm:loclization}. It  will be based on the
Wegner estimate shown in  \cite{EH2}
and a multi-scale analysis as used in \cite{EH2} to prove localization  at the bottom of the spectrum.
To this end, one needs an   initial length scale estimate. In \cite{EH2} an elementary
lower bound on the spectrum was  sufficient to obtain an initial length scale estimate for the bottom
of the spectrum.  This bound is not helpful for higher band edges and instead we use Theorem \ref{thm:boundonspec},
which provides an outer bound on the spectrum.
There is an additional difficulty originating from the fact that for the finite volume  Hamiltonians
there is always spectrum
 not only near the unperturbed Landau levels but also well in between them
 which corresponds to  states which live close
to the boundary. To deal with this difficulty, we use the Wegner estimate to estimate the exponential
decay of  the finite volume Hamiltonian in terms of the exponential decay of the infinite volume Hamiltonian,
this is outlined in  Subsection \ref{sec:wegdecayboxhamilonians}.
In Subsection \ref{sec:initiallength} we show the  initial length scale estimate and in Subsection \ref{sec:multiscale} we introduce
the multi-scale analysis, which is used to prove Theorem  \ref{thm:loclization}.

\section{Outer bound on the spectrum } \label{sec:outerbound}

Let $B(x) = B_0 + B_1(x)$, where $B_0$ is a constant magnetic field and $B_1$ denotes a
 non-constant  perturbation.
We define
\be\label{Esup}
e_{n,{\rm min}}[B] := \inf_{x \in \R^2} \left[ (2 n + 1 ) B(x) + V(x) \right] , \quad
e_{n,{\rm max}}[B] := \sup_{x \in \R^2} \left[ ( 2 n + 1 ) B(x) + V(x) \right].
\ee

These numbers correspond to the maximal broadening of
the Landau levels $(2n+1)B_0$ for the constant $B_0$ field if
the perturbation given by $B_1$ and $V$ were considered by classical mechanics.

\medskip

\begin{remark}
Note that
\be \label{eq:equivdef}
E_{n}^{+} = e_{n,{\rm max}}[B_{\omega_+}] , \quad E_{n}^{-} = e_{n,{\rm min}}[B_{\omega_-}] .
\ee
 \end{remark}

\medskip

To formulate the next theorem, we introduce the following quantities,
\begin{align*}
\hat{K}_2 &:= \| \nabla B_1 \|_\infty + \| \nabla V \|_\infty  \\
\hat{K}_3 &:= \| \nabla B_1 \|_\infty^2 .
\end{align*}

The following  theorem shows that in the large $B_0$ regime
the classical edges $e_{n,{\rm min}}[B]$ and $e_{n,{\rm max}}[B]$
give an outer bound on the true spectrum up to a small correction.

\begin{theorem} \label{thm:boundonspec} Let $C_1 > 0$ and assume that
\begin{align}
\label{eq:simeq}
  C_1^{-1} B_0  \leq B \leq C_1 B_0 .
\end{align}
Then for every $n \in \N$
 there exists a  constant $C_{{\rm ext},n} \geq 0$ (depending only on $C_1$)
  such that
the intervals $\hat{I}_n[B] := \left(\hat{I}_n^+[B], \hat{I}_{n+1}^-[B] \right) $ with
\begin{align} \label{eq:intofres}
\hat{I}_n^+[B] &:=    e_{n,{\rm max }}[B]  +
C_{{\rm ext},n} [ \hat{K}_2   B_0^{-1/2}  + \hat{K}_3 B_0^{-2} ]  , \\
\hat{I}_{n+1}^-[B] &:=  e_{n+1,{\rm min}}[B] -
C_{{\rm ext},n+1}  [ \hat{K}_2   B_0^{-1/2}  + \hat{K}_3 B_0^{-2} ]   , \label{eq:intofres2}
\end{align}
are in the resolvent set of $H(B)$.
\end{theorem}

\medskip

To prove the theorem we localize the resolvent in position space and treat it as a
 perturbation of the resolvent of  a Hamiltonian
with a constant magnetic field.  The choice of the  constant magnetic field
 will depend on  specific location in position space.

\medskip

\noindent
\begin{proof} We fix a gauge $A$ such that $\nabla\times A=B$.
To localize in position space,
we choose  a  profile function,  which equals the   normalized characteristic function  of a disk, centered at the origin, with radius one,
$g(x) := \frac{1}{\pi  } {\bf 1}_{\{ |x| < 1 \}}$. We use this function  to define
a rescaled and translated version thereof, by setting    $g_u(x) := g_{u,B_0}(x)  := B_0 g ( B_0^{1/2}(x-u) )$, with $u \in \R^2$.
The prefactor $B_0$ is chosen such that
\be \label{eq:normalgu}  \int g_u(x) du = 1 .\ee
  Moreover, we will denote the characteristic
function of the support of $g_u$ by  $\chi_{u} := \chi_{u,B_0} : =  {\bf 1}_{{\rm supp} g_{u}}$.

Next we introduce suitable gauges to control  the magnetic field on the support of $g_u$.
For each $u \in \R^2$ we set
$$
   B_{u,0} := B(u), \quad \mbox{and} \quad \delta_u[ B] := B - B_{u,0},
$$
 and
  we define the vector potentials
\begin{align}
\wt{A}_{u,0}(x_1,x_2)  &:  = \Big( - \frac{B_{u,0}}{2}(x_2-u_2) , \frac{B_{u,0}}{2}(x_1-u_1)
\Big) \label{defofnablalambdau2} \\
 \label{deltaAA}
\delta_u[\wt{A}](x_1,x_2)  &:= \Big(-\frac{1}{2} \int_{u_2}^{x_2}
 \delta_u[ B](x_1 ,\xi ) d \xi , \quad \frac{1}{2} \int_{u_1}^{x_1} \delta_u[  B ](\xi , x_2 ) d \xi \Big)  \\
\wt{A}_u  &:= \wt{A}_{u,0} + \delta_u[ \wt{A} ]  \label{defofnablalambdau} .
\end{align}
As a consequence of the definition we have
$
 \nabla \times \wt{A}_u  = B
$.
Thus if we define  the function
$$
\lambda_u(x) = \int_{\gamma_{x}}  ( \wt{A}_u  - A ) d \gamma_{x} ,
$$
where $ \gamma_{x}$ denotes any differentiable path connecting the origin with  $x \in \R^2$
(for example a straight line), then
\begin{align*}
& A = \wt{A}_u - \nabla \lambda_u .
\end{align*}
The function $\lambda_u$ will be used below in \eqref{eq:reseq}  for a gauge transformation,  and  we set
\begin{align}
A_{u,0} &:= \wt{A}_{u,0}  -  \nabla \lambda_u .
\end{align}
As an immediate consequence of the definition  we have
\be
\label{eq:constgauge} \nabla \times A_{u,0} = B_{u,0}  .
\ee

Moreover, we introduce the notation  $V_u = V(u)$, $\delta_u [ V ] = V- V_u$,  and for any vector potential $a$ we write $H_u(a) = (p - a)^2 + V_u $.

We consider   the following  resolvent identity which depends on the position $u$,
\begin{eqnarray*}
\frac{1}{z - H(A) } g_u  = \frac{1}{z - H_u(A_{u,0})} g_u  + \frac{1}{z - H(A)}  \big[ H(A) - H_u(A_{u,0})  \big] \frac{1}{z-H_u(A_{u,0}) } g_u
\end{eqnarray*}
which we integrate over $u$ as a weak integral and obtain, using  the above gauge transformation,
\be \label{eq:reseq}
\frac{1}{z - H(A) }  = \int \frac{1}{z - H_u(A_{u,0}) } g_u  du + \frac{1}{z - H(A)}  \int  e^{- i \lambda_u}  S_{u,z}  e^{ i \lambda_u} g_u du,
\ee
where we defined
\be \label{eq:introofgauge}
S_{u,z} := \big[H(\wt{A}_{u}) -  H_u(\wt{A}_{u,0})  \big] \frac{1}{z-H_u(\wt{A}_{u,0} )}  \chi_u  .
\ee
Now by   \eqref{eq:reseq} and the triangle inequality
\be \label{boundonerror0} \| (z - H(A) )^{-1} \| \leq \left\|   \int \frac{1}{z - H_u(A_{u,0}) } g_u  du  \right\| +  K \| (z - H(A))^{-1} \|,
\ee
where
\be \label{boundonerror}
K   := \left\|   \int e^{- i \lambda_u} S_{u,z}  e^{ i \lambda_u} g_u du  \right\| .
\ee
By  Proposition   \ref{prop11}, below, the first term on the right hand side of \eqref{boundonerror0} is bounded.
Thus  $(z - H(A) )^{-1}$ will be bounded and the theorem will follow, provided we show that $K < 1$.
But this follows from Lemma \ref{trivialineq} and  Lemma \ref{eq:kernelbound}
provided  $C_{{\rm ext},n}$ is chosen sufficiently large. \end{proof}

\medskip

\begin{lemma} \label{trivialineq} For $u \in \R^2$ let  $T_u$ be an operator in  $L^2(\R^2)$ with integral kernel $T_u(x,y)$ depending measurably
on $u,x,y$. Let $C$ be a bounded operator with integral kernel satisfying $ | T_u(x,y) | \leq  C(x,y)  $ for all $u\in \R^2$. Then
$$
\left\| \int T_u g_u du  \right\| \leq \| C \| .
$$
\end{lemma}

\medskip

\noindent
\begin{proof} Let $\psi \in L^2(\R^2)$. Then using \eqref{eq:normalgu} we find
$$
\left| \left[ \int T_u g_u du \psi\right](x)  \right|
\leq \int C(x,y) g_u(y) |  \psi(y)|   dy du  = [ C | \psi |](x) .
$$
Thus
$$
\left\| \int T_u g_u du \psi   \right\|  \leq  \|  C | \psi | \|  \leq \| C  \| \| \psi \| .
$$
\end{proof}

\medskip

\begin{proposition} \label{prop11}  Let $z \in \C$ with
${\rm dist}({\rm Re} z , \bigcup_{n \in \N_0 }  \left[e_{n,{\rm min}}[B] , e_{n,{\rm max}}[B] \right] ) > 0$
and $| {\rm Im} z | \leq B_0$. Suppose \eqref{eq:simeq} holds. Then
$$
\int du  \frac{1}{z - H_u(A_{u,0}) } g_u
$$
is bounded.
\end{proposition}

\medskip

\noindent
\begin{proof}  Recall that  $A_{u,0}$ is a vector potential for the constant magnetic field $B_{u,0}$,  \eqref{eq:constgauge}. Thus by definition  \eqref{Esup}
it follows that $\sigma( H_u(A_{u,0}) ) \subset \bigcup_{n \in \N_0 }  \left[e_{n,{\rm min}}[B] , e_{n,{\rm max}}[B] \right]$.
Now we  use Lemma \ref{lem:korpus} in the appendix, to estimate the integral kernel of the resolvent.
We conclude using \eqref{eq:boundongamma}, \eqref{eq:boundonU}, and \eqref{eq:simeq}
 that there exist an operator $D_z$ with integral kernel $D_z(x,y)$ such that for all $u \in \R^2$
$$
 \left| \frac{1}{z - H_u(A_{u,0})}(x,y) \right| \leq D_z(x,y) ,
$$
with
$$
D_z(x,y) = {\cal D}_z(x-y) ,  \quad   {\cal D}_z(x) \leq C_z ( 1  + \ln |x| ) \exp(-c_z |x|^2),
$$
for some constants $C_z$  and $c_z > 0$.
From this it follows that  $D_z$ is a bounded operator (with bound depending on $B_0$), since
 $\| {\cal D}_z \|_1$ is bounded and hence also the  Fourier transform of ${\cal D}_z$.
Now the claim follows in view of  Lemma \ref{trivialineq}.
\end{proof}

\medskip

The next lemma estimates the operator $S_{u,z}$ for $z$ in the following
neighborhood of the $n$-th Landau level,
\be \label{eq:neighlandau}
{\rm Re } z \in B_0[2n,2n+2] , \quad |{\rm Im} z | \leq B_0 .
\ee

\begin{lemma} \label{eq:kernelbound} Fix $n \in \N_0$ and let $z \in \C$ satisfy \eqref{eq:neighlandau}.
Suppose \eqref{eq:simeq} holds and let  $\sigma_n := \left[ e_{n,{\rm min}}[B]  ,   e_{n,{\rm max}}[B] \right]$. Then
\begin{align*}
| S_{u,z}(x,y) |
\leq &  \frac{ \hat{K}_2 B_0^{-1/2} + \hat{K}_3 B_0^{-2}     }{{\rm dist}({\rm Re} z , \sigma_n  )}Q_n(x,y)        ,
\end{align*}
where
\begin{equation} \label{kernelQ}  Q_n(x,y) :=  {\cal Q}_n(x-y),  \quad  {\cal Q}_n(x) :=     B_0    P_n(B_0^{1/2} | x| ) {\rm exp}\Big( - \frac{1}{8} B_0 |x|^2) ,
\end{equation}
and     $P_n(x) = C_n ( 1  + |x|^{-1} + |x|^{2(n+1)}  ) $ for some  $C_n$.  The  operator $Q_n$  is bounded uniformly  in $B_0$.
\end{lemma}

\medskip

\noindent
\begin{proof} The last sentence is a
consequence of \eqref{kernelQ}, which can be seen
by observing that $\| {\cal Q}_n \|_1$ and hence the  Fourier transform of ${\cal Q}_n$
 are uniformly bounded in $B_0$. In the proof we shall write $C$ for a numerical constant
 which may depend on $n$.
For notational simplicity we set
$$
\Pi_u = (- i \nabla - \wt{A}_{u,0} ) .
$$
Then
$$
H(  \wt{A}_u) - H(  \wt{A}_{u,0})   =  (  \delta_u[ \wt{A}])^2 +  i \nabla \cdot \delta_u[\wt{A}]  -  2 \delta_u[ \wt{A}] \cdot \Pi_u  + \delta_u[ V ] .
$$
Thus we find
\begin{align}
S_{u}   =  ( \delta_u[ \wt{A}] )^2 R_{u}  \chi_{u,B_0 }  + i ( \nabla \cdot \delta_u[ \wt{A}] )  R_{u}  \chi_{u,B_0 }  -  2 \delta_u[\wt{A} ]  \cdot \Pi_u R_{u}
\chi_{u,B_0 } + \delta_u [ V ]  R_{u}  \chi_{u,B_0 } \label{eq:expanofS} ,
\end{align}
where we defined $R_{u}  := (z - H_u(A_{u,0}) )^{-1}$ and  for simplicity we omitted $z$ from the notation.
To estimate the right hand side of \eqref{eq:expanofS}  note first that from \eqref{deltaAA}  one has the bounds
\begin{align}
& | \delta_u[\wt{A}](x) |  \leq C \| \nabla B_1 \|_\infty | x - u |^2  \label{expdeltaA1} \\
& | \nabla \cdot \delta_u[ \wt{A}](x) |  \leq C \| \nabla B_1  \|_\infty | x - u | \label{expdeltaA2} \\
& | \delta_u[ V ] |   \leq C \| \nabla V \|_\infty | x- u|\label{expdeltaA3} .
\end{align}
Moreover, we will use the explicit expression of the integral kernel for
the resolvent,
$$
R_{u}(x,y)  = \frac{1}{4 \pi} \left[   \Gamma_{u} U_{u} E_{u} \right](x,y) ,
$$
as given in   Lemma \ref{lem:korpus}, where, with  $z_u: = z - V_u$,
we define
\begin{align*}
\Gamma_{u}(x,y) &:=  \Gamma\Big(\frac{1}{2} - \frac{z_u}{2 B_{u,0} } \Big)   , \quad
U_{u}(x,y) :=  U\Big(\frac{1}{2}  - \frac{z_u}{2 B_{u,0} } ,\; 1 ;\; \frac{B_{u,0}}{2}|x-y|^2\Big) , \\
E_{u}(x,y) &= \exp\Big( - \frac{B_{u,0}}{4}|x-y|^2 - i \frac{B_{u,0}}{2}[x - u,y - u] \Big) .
\end{align*}
Here $\Gamma(\cdot)$ stands for the usual Gamma-function
and $U(\cdot, \cdot ; \cdot)$ is the confluent hypergeometric
function. For more details, see Appendix~\ref{appres}.

Using this representation,
  one  finds   for the two components $([\Pi_u]_1 R_u, [\Pi_u]_2  R_u)$
 of $\Pi_u R_u$ that
$$
[\Pi_u]_{1,2} R_{u}(x,y) = \left[ i \frac{B_{u,0}}{2}(x-y)_{1,2} + (-1)^{2,1} \frac{B_{u,0}}{2}(x-y)_{2,1}
 \right] R_{u}(x,y)   - i B_{u,0} (x - y)_{1,2} R_{u}'(x,y),
$$
where we introduced $R_u' :=  \frac{1}{4 \pi} \Gamma_{u} U_{u}' E_{u}$ with
$
U_u'(x,y) =  U_3\left(\frac{1}{2}  - \frac{z_u}{2 B_{u,0} } , \, 1 ; \, \frac{B_{u,0}}{2}|x-y|^2\right)
$
($U_3$ denotes  the partial derivative with respect to the third variable).
We conclude  that
\be \label{eq:boundonpiR}
 | (\Pi_i R_u)(x,y)|\le C B_{u,0} |x-y| \left[ | R_u(x,y)|  + | R_u'(x,y) | \right]  , \quad i=1,2 .
\ee
Now using \eqref{eq:boundonpiR}, the bounds \eqref{expdeltaA1}--\eqref{expdeltaA3}, the triangle inequality
  $| x - u | \leq | x - y | + | y - u |$, and
  $\chi_{u,B_0 }(y) | y - u | \leq C B_0^{-1/2}$, to estimate \eqref{eq:expanofS} we obtain
\begin{align}
| S_{u}(x,y) |  \label{suest1} \leq
&   \;  C \| \nabla B_1 \|_\infty^2  \Big[    B_{u,0}^{-2}  \{ B_{u,0} |x-y|^2 \}^2 |R_{u}(x,y)|   +    B_0^{-2}   |R_{u}(x,y)|  \Big] \chi_{u,B_0 }(y)        \\
 &  +  C \| \nabla B_1  \|_\infty  \Bigg[  B_{u,0}^{-1/2}  \{ B_{u,0} |x-y|^2 \}^{1/2} |R_{u}(x,y)|  +    B_0^{-1/2}   |R_{u}(x,y)|  \nonumber      \\
  & +  B_{u,0}^{-1/2}    \{  B_{u,0} |x-y |^2 \}^{3/2} \left( |R_{u}(x,y)|   +    |R_{u}'(x,y)|           \right) \nonumber    \\
&   +  B_{u,0}^{1/2}  B_{0}^{-1}    \{ B_{u,0} |x-y |^2 \}^{1/2} \left( |R_{u}(x,y)|
+  |R_{u}'(x,y)|            \right)   \Bigg] \chi_{u,B_0 }(y)  \nonumber \\
&   +  C  \| \nabla V \|_\infty \Big[ B_{u,0}^{-1/2} ( B_{u,0} | x - y |^2)^{1/2} +   B_0^{-1/2} \Big]     |R_{u}(x,y)|             \chi_{u,B_0 }(y). \nonumber
\end{align}
Now using  \eqref{eq:boundongamma}, \eqref{eq:boundonU},  \eqref{eq:boundonU2}, and \eqref{eq:simeq}
to estimate  \eqref{suest1},  we find for $z$ satisfying \eqref{eq:neighlandau},
\begin{eqnarray*}
 | S_{u}(x,y) |  & \leq
&  \frac{C  [ \hat{K}_2 B_0^{-1/2}  + \hat{K}_3 B_0^{-2} ]}{{\rm dist}({\rm Re} z , \sigma_n  )}
  B_{u,0} P_n(B_{u,0}^{1/2} | x - y | ) {\rm exp}\Big( - \frac{1}{4} B_{u,0} |x-y|^2\Big)
\chi_{u,B_0 }(y) ,
\end{eqnarray*}
with $P_n$ of the  form as stated in the lemma.
  Now in view of \eqref{eq:simeq} the bound in the lemma now follows.
\end{proof}

\section{Location  of the Spectrum}

\label{sec:detspec}

In  Subsection
\ref{sec:detspecb}  we prove Theorems  \ref{thm:estonspeca}  and \ref{thm:estonspecb}.
To this end we  first derive in Subsection  \ref{sec:detspeca}      two deterministic results Lemma \ref{lem:parta}
and  \ref{lem:partb}

\subsection{Deterministic Part}
\label{sec:detspeca}
The following  preparatory Lemma is a trivial consequence of the spectral theorem.

\begin{lemma} \label{lem:spec} Let $H$ be a self-adjoint operator $E \in \R$ and
$\psi \in \mathcal{H}$, $\psi\ne 0$. Then
$$
\| ( H - E ) \psi \| \leq \epsilon \| \psi \| \Rightarrow \sigma(H) \cap [E- \epsilon , E + \epsilon ] \neq \emptyset .
$$
\end{lemma}

\medskip

\noindent
\begin{proof} Suppose $\sigma(H) \cap [E- \epsilon , E + \epsilon ] =  \emptyset$. Then by the spectral theorem the assumption implies
$$
\| \psi \| = \| (H - E)^{-1} (H-E) \psi \| < \frac{1}{\epsilon} \epsilon \| \psi \| ,
$$
which is a contradiction.
\end{proof}

\medskip

The following Lemma holds for an arbitrary magnetic field $B$, which is bounded from below.
It will be used in the proof of  Theorem  \ref{thm:estonspeca}.

\medskip

\begin{lemma} \label{lem:parta}  Let $B(x)$ be a magnetic field
with
$$
B_{\rm inf} := {\rm inf}_x B(x) > 0.
$$
Then for any $n \in \N_0$  there exists a constant $C_n$ (depending on $n$) such that
for all $\lambda$ in the range of the function $(2n+1)B + V$,
$$
{\rm dist}(\sigma(H(B)),\lambda) \leq C_n
\left( \big[  \| \nabla B \|_\infty + \| \nabla V \|_\infty \big]
  B_{\rm inf}^{-1/2}  +  \| \nabla B \|_\infty^2  B_{\rm inf}^{-2} \right)   .
$$
\end{lemma}

\noindent
\begin{proof}
By assumption, there exists an $\widetilde{x} \in \R^2$ such that
$\lambda  = V(\widetilde{x})   + ( 2n+1) B(\widetilde{x}) $.
Choose  a  gauge
$$
A(x)  = \frac{1}{2} \left(  - \int_{\widetilde{x}_2}^{x_2} B(x_1, y_2) \rd y_2 ,
   \int_{\widetilde{x}_1}^{x_1} B(y_1, x_2) \rd y_1        \right) ,
$$
and  set $A_{0}(x)  := \frac{1}{2}B_{0}( -(x_2 - \widetilde{x}_2), x_1- \widetilde{x}_1 )$
with $\wt{B}_{0} :=  B(\widetilde{x})$.
Let us consider the normalized trial state
\be
\varphi_n(x) = \left\{ \frac{\wt{B}_{0} }{2 \pi} \right\}^{1/2} L_{n}(\wt{B}_{0}|x -
\widetilde{x}|^2 / 2 )   \exp(-\frac{1}{4}\wt{B}_{0}|x - \widetilde{x}|^2  ) ,
\ee
where $L_n$ is the $n$-th Laguerre polynomial.
We set
$$
\delta [A]  := A  - A_0 , \quad \delta [ V ] := V - V(\wt{x}) .
$$
Expanding the square, one finds
\begin{align}
H( A ) \varphi_n &= H( A_0) \varphi_n + [ H( A ) - H( A_0) ] \varphi_n \nonumber \\
& = \lambda \varphi_n + \{  \delta [ V ] + (\delta [A])^2 +
i \nabla  \cdot \delta [ A ] - 2 \delta [ A ] \cdot \wt{\Pi} \}
 \varphi_n , \label{eq:expsquare1}
\end{align}
with  $\wt{\Pi} := (p-A_0)$. Using the estimates
\begin{align*}
| \delta[ A](x) | & \leq C \| \nabla B \|_\infty | x - \tilde{x} |^2   \\
| \nabla \cdot \delta[A] (x) | & \leq C \| \nabla  B \|_\infty | x - \tilde{x} |  \\
| \delta [ V  ] | & \leq \| \nabla V \|_\infty | x - \wt{x} |  ,
\end{align*}
and \eqref{eq:expsquare1}, we find that
\be  \label{eq:trialest3}
\| H(A) \varphi_n -  \lambda \varphi_n \|
\leq C \left[ \frac{ \| \nabla B \|_\infty + \| \nabla V \|_\infty }{\wt{B}_0^{1/2} } +  \frac{\| \nabla B \|_\infty^2}{\wt{B}_0^2 } \right]  .
\ee
The lemma now follows in view of  Lemma \ref{lem:spec} and since by definition $\wt{B}_0 \geq B_{\rm inf}$.
\end{proof}

\medskip

The following Lemma estimates the change of the  spectrum of a  magnetic Schr\"odinger with arbitrary
magnetic field $B$  under a the perturbation by a small magnetic field $B'$.
It will be used in the proof of  Theorem  \ref{thm:estonspecb}.

\medskip

\begin{lemma} \label{lem:partb}
 Let $\epsilon > 0$, $K >1$, and $v>0$. Then there exists an $\eta > 0$
such that the following holds.
For any magnetic field $B$, potential $V$, with $\| V \|_\infty \leq v$, energy
 $E \in \sigma(H(B))$, with $E + 1 \leq K$, and
magnetic field  $B'$, with
 $\| B' \| + \| \nabla B' \| \leq \eta$, one has
  $$[ E- \epsilon , E + \epsilon] \cap  \sigma(H(B + B') ) \neq \emptyset .$$
\end{lemma}

 \medskip

The crucial part of the lemma is that the $\eta$ does not depend on the magnetic  field $B$.

 \medskip

\noindent
\begin{proof}
First, we will show that there exists a trial  state which is localized in a box of finite side
length $L$ (depending only on $\epsilon$, $K$, and $v$), see  \eqref{ineq:weyl44}.
Then we can use perturbation theory to complete the proof.

We consider the following partition of $\R^2$.
Let $\chi$ be a smooth function with support contained in $B_{2}(0) = \{ x \in \R^2 : |x| \leq 2 \} $
such that $\sum_{z \in \Z^2}  \chi^2_z(x) = 1$, where $\chi_z(x) = \chi(x-z)$. We assume
that at most four different $\chi_z$'s overlap, that is
\be \label{eq:overlapless4}
\sum_z 1_{{\rm supp} \chi_z} \leq 4 .
\ee
Set  $\chi_{z,L}(x) = \chi(L^{-1}(x-z))$. Without loss we can assume that  $0 < \epsilon \leq 1$.
 Let $\psi$ be a normalized state such that
\be \label{eq:initialtrial}
   \| ( H(A) - E) \psi \| \leq \frac{\epsilon}{100}  .
\ee
Choose $L$ sufficiently large such that
\be \label{eq:Lepsilon}
\frac{ 80 \| \nabla \chi \|_\infty^2}{L^2} ( K  +  v ) \leq \frac{\epsilon^2}{100}    ,      \quad
 \frac{ 40 \| \Delta \chi \|_\infty^2}{L^4}  \leq \frac{\epsilon^2}{100}    .
\ee
We claim, that there exists a $z \in \Z^2$ such that
\begin{align}
 \| \chi_{L,z} \psi \|^2  &\geq    \frac{L^2}{80 \| \nabla \chi \|_{\infty}^2  ( K  + v )}
 \| \nabla \chi_{L,z} \cdot \Pi \psi \|^2  \label{eq:A} \\
  \| \chi_{L,z} \psi \|^2  &\geq    \frac{L^4}{40 \| \Delta  \chi \|_{\infty}^2  }
  \| \Delta  \chi_{L,z} \psi \|^2  \label{eq:AA} \\
\| \chi_{L,z} \psi \|^2 &\geq \frac{100}{\epsilon^2} \|\chi_{L,z} \xi \|^2
 \label{eq:B} , \end{align}
where $\xi :=  ( H - E) \psi $ and $\Pi := p - A$. Suppose this were not the case: then for all $z \in \Z^2$
 one of the inequalities \eqref{eq:A}--\eqref{eq:B}
would not hold and this would imply the first inequality of the following estimate
\begin{align}
1 &= \sum_{z \in \Z^2} \| \chi_{L,z} \psi \|^2  \nonumber \\& \leq
\frac{L^2 }{80 \| \nabla \chi \|_{\infty}^2 (K + v ) }
\sum_{z} \| \nabla \chi_{L,z} \cdot \Pi \psi \|^2  +  \frac{L^4 }{40 \| \Delta
\chi \|_{\infty}^2  } \sum_{z} \|  \Delta \chi_{L,z}  \psi    \|^2
     + \frac{100}{\epsilon^2} \sum_{z}  \| \chi_{L,z} \xi \|^2  \nonumber \\
     & \leq \frac{4}{10} , \label{eq:contra}
\end{align}
where  the first term on the second line were estimated by $1/10$ as follows,
\begin{align*}
\sum_z \| \nabla \chi_{L,z} \cdot \Pi \psi \|^2  & \leq
       \sum_z 2  \sum_{i=1}^2 ( \Pi_i \psi ,   (\nabla_i \chi_{L,z} )^2   \Pi_i \psi   )    \\
& \leq      8 \| \nabla  \chi  \|_\infty^2  L^{-2}  \sum_{i=1}^2 ( \Pi_i \psi ,   \Pi_i \psi   )
=   8 \| \nabla \chi  \|_\infty^2  L^{-2} [ ( \psi ,  H(A) \psi)  + v ]  \\
& \leq  8 \| \nabla \chi  \|_\infty^2  L^{-2}  ( K + v ) ,
\end{align*}
 where we used  \eqref{eq:overlapless4} in the second line and \eqref{eq:initialtrial} in the last.
 The other terms in the second line of \eqref{eq:contra}
 can be estimated in a similar but easier way.
But  \eqref{eq:contra}  yields a contradiction.  Thus let $z^* \in \Z^2$ be such that
\eqref{eq:A}--\eqref{eq:B} hold. Then  calculating a commutator, using the triangle inequality, we find
\begin{align}             \nonumber
  \| (H(A) - E) \chi_{L,  z^*} \psi \|  &=  \|    \chi_{L,  z^*} \xi    - 2 i \nabla \chi_{L,  z^*} \cdot \Pi \psi  - \Delta \chi_{L,  z^*} \psi   \|             \\
& \leq \| \chi_{L,z^*} \xi \| + 2 \| \nabla \chi_{L,z^*} \cdot \Pi \psi \|  +   \| \Delta \chi_{L,  z^*} \psi   \|    \leq  \frac{\epsilon}{2} \| \chi_{L,  z^*} \psi   \| ,
   \label{ineq:weyl44}
\end{align}
where the last inequality  is a consequence of  \eqref{eq:A}--\eqref{eq:B} and the choice of $L$, \eqref{eq:Lepsilon}.

Now we can use ordinary perturbation theory. We choose the gauge
\begin{equation} \label{deltaA}
A'(x_1,x_2) := \Big(-\frac{1}{2} \int_{z^*_2}^{x_2} B'(x_1 ,y ) d y ,\;
 \frac{1}{2} \int_{z^*_1}^{x_1} B'( y , x_2 ) d y \Big) .
\end{equation}
Let $\varphi = \chi_{L,  z^*} \psi$ and let $G$ denote  the support of $\chi_{L,  z^*}$. Then by \eqref{deltaA} we have
\be \label{eq:growthofA}
\| (A')_j {\bf 1}_G \|_\infty \leq  \| B' \|_\infty  L , \quad
\| (A')^2 {\bf 1}_G \|_\infty \leq 2 \| B' \|_\infty^2   L^2, \quad    \| \nabla \cdot A'  {\bf 1}_G \|_\infty \leq  ( \| B' \|_\infty   +  2 \| \nabla B' \|_\infty  L ) .
\ee
We find,  using first the triangle inequality and then  \eqref{ineq:weyl44} and  \eqref{eq:growthofA},
\begin{eqnarray}
\lefteqn{ \| [(p - A - A' )^2 + V - E ] \varphi \|  } \nonumber \\
&& \leq   \| [ ( p - A)^2  + V - E ]  \varphi \| +  \| [ A'^2 + i \nabla \cdot A']\varphi \|  + \|  2  A' \cdot (p-A)  \varphi \|
\nonumber  \\
 &&  \leq
\Big[ \frac{\epsilon}{2}  +   2 \| B' \|_\infty^2 L^2 +  \| B' \|_\infty   +
2 \| \nabla B' \|_\infty  L  \Big] \| \varphi \|   +  2  \| B' \|_\infty  L  \Big[\| (p-A)_1 \varphi \| + \| (p-A)_2 \varphi \| \Big]. \label{eq:perturbest3}
\end{eqnarray}
Now inserting the estimate
$$
   \Big[\| (p-A)_1 \varphi \| + \| (p-A)_2 \varphi \| \Big]^2 \leq 2 ( \varphi , (p-A)^2 \varphi ) \leq 2 ( K + v )  \| \varphi \|^2 ,
$$
into \eqref{eq:perturbest3} we find
\begin{align*}
\| [(p - A - A' )^2 + V - E ] \varphi \|  \leq   \epsilon  \| \varphi \| ,
\end{align*}
provided we choose $\eta > 0$ sufficiently small (depending only on $K$ and $L$).
The Lemma now follows in view of Lemma   \ref{lem:spec}.
\end{proof}

\medskip

\subsection{Probabilistic Part}
\label{sec:detspecb}
To show  Theorems \ref{thm:estonspeca} and \ref{thm:estonspecb},
we will combine the previous two Lemmas concerning a deterministic
magnetic field, with the following probabilistic result \cite[Theorem 8.1]{EH2}.

\begin{theorem}   \label{thm:ibspec}  Suppose $B_\omega$ is a random magnetic
field constructed   in \eqref{B}, \eqref{betadef}, and \eqref{bconst},
satisfying \eqref{lowbound1} and {\bf (i.i.d.)} and $\rho > ln 2$. Assume $B_{\rm var}$ and
$V$ are $\Z^2$-periodic.  Then for all $\omega$ in the support of the probability measure we have
$$
\Sigma \supset  \sigma( H( B_{\omega})  )  .
$$
\end{theorem}

\medskip

\noindent
{\bf Proof of Theorem \ref{thm:estonspeca}}. The theorem
 follows since by  Theorem   \ref{thm:ibspec}  we have $\sigma(H(A_{\omega^\pm})) \subset \Sigma$ and  Lemma \ref{lem:parta}.
\qed

\medskip

\noindent
{\bf Proof of Theorem \ref{thm:estonspecb}}.   By Theorem
\ref{thm:deterministic}  there exists an $\omega^*$ such that 
$\Sigma = \sigma(H(A_{\omega^*}))$.
In particular, it follows that $\Sigma$ is a closed set.
 Let   $E \in \Sigma$ and let $\epsilon > 0$.
Now we consider the path $\omega_t = t \omega^*$, with $t \in [0,1]$.
Using Lemma  \ref{lem:partb}, we can find a sufficiently large $N$ 
such that for the numbers
$t_i = 1 - \frac{i}{N}$, with $i=1,...,N$, there exist  $E_i \in \sigma(H_{\omega_{t_i}})$  satisfying $|E_i - E_{i-1}| \leq \epsilon$.
Clearly, $E_N \in \{ B_0 (2n + 1 ) \; : \; n \in \N \}$.
By Theorem  \ref{thm:ibspec}, we know that $E_i \in \Sigma$. Since
  $\epsilon >0$ can be
chosen arbitrarily small by choosing $N$ sufficiently large,
it follows that  all numbers between $E$ and $B_0 (2n+1)$, for some $n \in \N_0$,
are contained in $\Sigma$,  since $\Sigma$ is closed.

 This implies the existence of the intervals as stated in the theorem.
It remains to show the estimate regarding the endpoints of $\Sigma_n$.
We will set 
\be\label{Cndef}
C_n = {\rm max}(C_{{\rm int},n}, C_{{\rm ext},n}).
\ee
Fix $n$. By  Theorem      \ref{thm:boundonspec} we can choose $B_0$ 
sufficiently large such that there is a gap in
the spectrum $\Sigma$
 located between the Landau levels $B_0(2n+1)$ and $B_0(2n+3)$ and that $E_{{\rm min},n}$ is to the left of the gap
and $E_{{\rm max},n+1}$ is to the right of the gap.
The estimate  \eqref{eq:estonbandedge}
regarding the endpoints of $\Sigma_n$ can be rephrased as  two inequalities.
One of the  inequalities  follows  in view of \eqref{eq:equivdef} from
Theorem \ref{thm:estonspeca}  and the other
 from Theorem      \ref{thm:boundonspec}.
\qed

\section{Localization} \label{sec:loc}

In this section we show Theorem  \ref{thm:loclization} using multi-scale analysis
and the Wegner estimate from \cite{EH2}. We have to show the initial
length scale estimate, which is the content of Subsections \ref{sec:wegdecayboxhamilonians},  \ref{sec:initiallength},
and \ref{sec:multiscale}.

By  $\Lambda \subset \bR^2$ we denote   a square.     We will consider the
magnetic Schr\"odinger operator  with Dirichlet boundary conditions
on $\Lambda$, and denote it by
\be
H_\Lambda(A) =   (p - A)^2  + V .
\ee
We realize this as a self adjoint operator by means of  the Friedrichs extension.
We will work in the Hilbert space $L^2(\Lambda)$ and
  denote the scalar product by $\langle \,\cdot ,\cdot\, \rangle$
and the norm by $\|\,\cdot\,\|$. In particular we will work with the following squares.
For $l > 0$ and $x \in \R^2$ we denote   by
\be \label{eq:defoflambdaboxes}
 \Lambda_l(x) := \{ y \in \R^2 :   | y - x |_\infty < l/2 \}
\ee
the open square  centered at $x$  with sidelength $l$.

If  $l\in \bN$,  we will write
\begin{equation} \label{eq:defofdirichletham}
H_l(A)  = H_{\Lambda_l(0)}(A) .
\end{equation}
Boxes with sidelength $l \in 2 \N + 1$ and center $x \in \Z^2$
are called {\it suitable}.
For  suitable squares, we set
$$
\Lambda^{{\rm int}} := \Lambda_{l/3}(x) , \quad
\Lambda^{{\rm out}} := \Lambda_{l}(x) \setminus \Lambda_{l-2}(x) ,
$$
and we set  $\chi^{\rm int} = \chi_{\Lambda^{\rm int}}$ and $\chi^{\rm out} =
 \chi_{\Lambda^{\rm out}}$.

We introduce the constant $c_\delta$ to be the smallest integer such that
\begin{equation} \label{def:cdelta}
c_\delta \geq \left\{ \begin{array} {ll} \frac{3}{2}  , & {\rm in \ \ case } \ \  \eqref{kcond} \\
                  \delta^{-1}  ,  &    {\rm in \ \ case } \ \  \eqref{kcond1}
                   \end{array} \right.
\end{equation}
which gives the distance beyond which the random magnetic field is independent.
We define  $\widetilde{\Lambda} := \Lambda + [-c_\delta,c_\delta]^2$.

In this section we consider the random magnetic field
$B_\omega = B_{\rm det} + \mu B_{\rm ran}^\omega$,
as introduced in Section \ref{sec:field}.
Let $A_\omega$ be a vector potential
with $\nabla\times A_\omega =  B_{\rm det} + \mu B_{\rm ran}^\omega$.
We introduce  a random magnetic field subordinate to the square $\Lambda$,
\be
 \widetilde{B}_{\Lambda,\omega} :=   B_{\rm det} + \mu   \widetilde{B}_{{\rm ran},\Lambda}^\omega , \quad
        \widetilde{B}_{{\rm ran},\Lambda}^\omega :=   \sum_{k=0}^\infty
 \sum_{z \in \Lambda^{(k)} \cap \wt{\Lambda}}  B^{(k)}_z ;
\label{bconstresr}
\ee
for notation we refer the reader to  \eqref{bconst}. Informally speaking,
the random field \eqref{bconstresr}  is obtained by adding to  the deterministic
magnetic field  the random magnetic field generated only  by
the random variables living on the  square  $\widetilde{\Lambda}$.
Likewise, for  a random vector potential
$A_\omega$  generating the  magnetic field $B_\omega$,  we introduce a random vector potential
subordinated to the square  $\Lambda$, centered at $z$, by
\begin{align*}
&\wt{A}_{\Lambda,\omega}(x) := A_\omega(x)    -  \Big( -  \frac{1}{2}  \int_{z_2}^{x_2}
\delta [\widetilde{B}_{\Lambda,\omega}](x_1,\xi) d \xi  , \;
\frac{1}{2}  \int_{z_1}^{x_1} \delta [\widetilde{B}_{\Lambda,\omega}](\xi,x_2) d \xi \Big) \\
&\delta [\widetilde{B}_{\Lambda,\omega}] : =    B_\omega  - \widetilde{B}_{\Lambda,\omega} .
\end{align*}
Observe that   $\wt{A}_{\Lambda,\omega}$
is a vector potential
with magnetic field $\widetilde{B}_{\Lambda,\omega}$ such that
$\wt{A}_{\Lambda,\omega}  = A_\omega$   on $\Lambda$.

For  an operator $T$  in a Hilbert space we will denote
by $\rho(T)$ the resolvent set of $T$.

\subsection{The Wegner Estimate and Exponential Decay} \label{sec:wegdecayboxhamilonians}

In this subsection we first state the Wegner estimate from \cite{EH2}.
Fix an energy $E$ and a window of width $\eta\leq 1$ about $E$.
Let $\chi_{E,\eta}$ be the characteristic function of the interval
$[E-\eta/2, E + \eta/2]$.

\medskip

\begin{theorem} \cite[Theorem 3.1]{EH2} \label{thm:wegner} Let  {\bf ($\boldsymbol{\rm W}$)} hold with  $K_0 > 3$,
$\rho>\ln 2$ and
$0 < \delta \leq \delta_0$. Let  $K_1 \geq 1$.
Then
there  exist positive  constants $C_0=C_0(K_0,K_1)$, $C_1 = C_1(K_0,K_1)$,
and $L_0^*=L_0^*(K_0,K_1,\delta)$  such that
for any $0 < \kappa \leq 1$
$$
   \mathbb{E} \, \Tr \chi_{E,\eta}(H_l(A)) \leq C_0  \eta \mu^{-2}
l^{ C_1(\kappa^{-1}  + \rho)} ,
$$
for all  $E \in [ \frac{b_0}{2} ,  K_1 b_0]  $,  $0 < \eta \leq 1$, and  $l\ge L_0^* b_0^{\kappa}$.
\end{theorem}

\medskip
Now we use the Wegner estimate and the geometric resolvent identity to show
the following  Lemma. It will be used to show that  for two given  {\em independent}
squares (squares which are sufficiently far apart such that their Hamiltonians are independent)
with very high  probability  for at least one of the squares the exponential
decay of the finite volume Hamiltonian can be estimated in terms of the
infinite volume Hamiltonian. In precise terms,  we say a square
$\Lambda = \Lambda_l(x)$ is {\bf $(E, C_\infty,\alpha)$--balanced}
 if $E \in \rho( (H_{\Lambda}(A) )$ and
 \be \label{eq:wegnerexpdecay}
\left\| \chi^{\rm out} (H_{\Lambda}(A) - E)^{-1} \chi^{\rm int} \right\|   \leq ( 1 +  C_\infty l^\alpha )
 \left\| \chi^{\rm out} (H(\wt{A}_{\Lambda})  - E)^{-1} \chi^{\rm int} \right\| .
\ee

\begin{lemma} \label{lem:wegini} Let  {\bf ($\boldsymbol{\rm W}$)} hold with  $K_0 > 3$,
 and let  $K_1 \geq 1$ and $0 < \kappa \leq 1$. There exist  constants $C_0,C_1,L_0^*$ (the same as  in Theorem \ref{thm:wegner})
 and a constant $C_\infty = C_\infty(K_1)$, such that for any $\alpha > 0$,  subinterval
$J \subset [ \frac{b_0}{2} ,  K_1 b_0]$  and any $x , y \in \Z^2$
 with  $| x - y |_\infty \geq l + c_\delta$,
  \begin{eqnarray}   \label{prob}
 \mathbb{P}\Big( \forall E \in J \; , \;  \Lambda_l(x) \ {\rm  or} \ \Lambda_l(y) \ {\rm is} \  (E, b_0 C_\infty, \alpha){\rm-balanced} \ \Big)
     \geq 1 -  3 C_0^2  |J| \mu^{-4} l^{ 2 C_1(\kappa^{-1}  + \rho)}  l^{- \alpha} ,
\end{eqnarray}
provided $l\ge L_0^* b_0^{\kappa}$.\\
 \end{lemma}

\noindent
\begin{proof} To shorten notation we set  $\beta = C_1(\kappa^{-1}  + \rho)$ and $D_0 =  C_0  \mu^{-2}$.

\medskip

\noindent
\underline{Step 1:} Let $E \in J$. Then we claim that with probability greater or equal than the
right hand side of
\eqref{prob} we have for $w = x$ or $w=y$ that
$ E \in \rho(H_{\Lambda(w)})$ and
\be \label{eq:weg}
\| (H_{\Lambda(w)}(A) - E)^{-1} \| \leq  8  l^\alpha .
\ee

\medskip

\noindent
To prove \eqref{eq:weg},
let $a$ denote the left endpoint of the interval $J$ and we
consider a partition of $J$ with respect to the points
$$
x_j = a + j l^{-\alpha} .
$$
Define the intervals
$
J_{j,1} =[x_{2j-1}, x_{2j+1}] $, and  $J_{j,2} =[x_{2j}, x_{2j+2}] $.
Then
\be \label{eq:star11}
J \subset \bigcup_{j=0}^{N} ( J_{j,1} \cup J_{j,2} ) ,
\ee
where $N$ denotes the smallest integer larger than $|J | l^\alpha/2$.
Using the independence of spectral properties
of the  local Hamiltonians in $\Lambda(x)$ and $\Lambda(y)$
in addition to the Wegner estimate, we find
$$
{\mathbb{P}}\{ \forall_{w \in \{x,y \}} J_{j,\sigma} \cap \sigma(H_{\Lambda(w)}) \neq \emptyset \}  =
 \prod_{w \in \{x,y \}} {\mathbb{P}}\{  J_{j,\sigma} \cap \sigma(H_{\Lambda(w)}) \neq \emptyset \}   \leq D_0^2 l^{-2 \alpha + 2 \beta } .
$$
Since the covering of $J$ given in \eqref{eq:star11} contains $2 N  (\leq 3 |J| l^\alpha )$ intervals, we have
$$
{\mathbb{P}}\{  \exists_{j  \in \{1,2,..,N\}} \exists_{\sigma \in \{1,2 \}}
\forall_{w \in \{x,y \}} J_{j,\sigma} \cap \sigma(H_{\Lambda(w)}) \neq \emptyset \} \leq 3 |J | D_0^2 l^{- \alpha + 2 \beta } .
$$
Thus the probability that this event does not occur can be estimated from below,
\be \label{prob22}
{\mathbb{P}}\{  \forall_{j  \in \{1,2,..,N\}} \forall_{\sigma \in \{1,2 \}}
\exists_{w \in \{x,y \}} J_{j,\sigma} \cap \sigma(H_{\Lambda(w)}) =
\emptyset \} \geq  1 - 3 |J| D_0^2 l^{- \alpha + 2 \beta }
\ee
By \eqref{eq:star11} for any  $E \in J$ there exists an interval $J_{j,\sigma}$
which contains $E$ such that the distance of $E$ to the boundary of $J_{j,\sigma}$ is
greater than $l^{-\alpha}/8$.  This observation and \eqref{prob22} imply the claim in Step 1.

\medskip
\noindent
\underline{Step 2:}  Step 1 implies  \eqref{prob}.

\medskip

For $w = x$ or $w=y$ we write $\Lambda = {\Lambda(w)}$.
By the geometric resolvent identity, we have
\be \label{eq:geom0}
\chi^{\rm out} (H_{\Lambda}(A) - E )^{-1} \chi^{\rm int} = \chi^{\rm out} \phi (H(\wt{A}_\Lambda) -E)^{-1} \chi^{\rm int} + \chi^{\rm out}
 (H_{\Lambda}(A) - E )^{-1} W(\phi) (H(\wt{A}_\Lambda) -E)^{-1} \chi^{\rm int} ,
\ee
with  $W(\phi) = 2 \nabla \phi \cdot (p - \wt{A}_\Lambda) + \Delta \phi $.
We choose $\phi \in C_0^\infty(\R^2; [0,1])$ to be a function such that
$$
\phi = 1 \quad {\rm on } \quad \Lambda_{l-1} , \quad
\quad  \phi = 0 \quad {\rm on} \quad \mathbb{R}^2 \setminus \Lambda_{l-1/2} .
$$
To estimate  the second term in \eqref{eq:geom0} we use
\be  \label{eq:geom1}
\| (\Delta \phi) (H(\wt{A}_\Lambda) - E )^{-1} \chi^{\rm int} \|
\leq C \| \chi^{\rm out} (H(\wt{A}_\Lambda) - E)^{-1} \chi^{\rm int} \| ,
\ee
and  we use Lemma B.1 of \cite{EH2} with
$u = (H(\wt{A}_\Lambda) - E)^{-1} \chi^{\rm int}$,
$\wt{\Omega} = {\rm supp} | \nabla \phi |$,
$\Omega = \Lambda^{\rm out}$, yielding
\be \label{eq:geom2}
\| \nabla \phi \cdot (p - \wt{A}_\Lambda) (H(\wt{A}_\Lambda) - E)^{-1} \chi^{\rm int} \| \leq C ( 1 + |E| )
\| \chi^{\rm out} (H(\wt{A}_\Lambda) -E)^{-1} \chi^{\rm int} \|
\ee
using that $\chi^{\rm int}=0$ on $\Omega$.
Now using  \eqref{eq:geom1}, \eqref{eq:geom2} and \eqref{eq:weg} to estimate \eqref{eq:geom0} Step 2 and hence the lemma follows,
since $E \in [\frac{b_0}{2},K_1 b_0]$.
\end{proof}

\subsection{Lifshitz asymptotics}

 \label{sec:initiallength}

In this subsection we show a Lifshitz estimate for the probability
that the operator $H(\wt{A}_\Lambda)$ with a local magnetic field
has an eigenvalue beyond
the outer bound on the infinite volume operator.
This result will
 imply a spectral estimate in Corollary \ref{lem:probestoninfspec2}
which will be used in the next section to obtain the
 initial length scale estimate.
In the theorem below we will use similar notation as introduced in Theorem  \eqref{thm:boundonspec},
but we take the supremum over
the essential support of the probability measure.
That is we define  $I_n[B] := \left({I}_n^+[B], {I}_{n+1}^-[B] \right) $ with
\begin{align} \label{eq:intofresesup}
{I}_n^+[B] &:=    e_{n,{\rm max }}[B]  +
C_{{\rm ext},n} [ {K}_2   B_0^{-1/2}  + {K}_3 B_0^{-2} ]  , \\
{I}_{n+1}^-[B] &:=  e_{n+1,{\rm min}}[B] -
C_{{\rm ext},n+1}  [ {K}_2   B_0^{-1/2}  + {K}_3 B_0^{-2} ]   , \label{eq:intofres2esup}
\end{align}
with  $K_3 := {\rm ess \, sup}_\omega \| B_\omega \|_\infty^2$.

\medskip

\begin{theorem}          \label{lem:probestoninfspec}
Assume that {\bf ($\boldsymbol{{\rm A}_\tau}$)} holds.
 Then for  $h  > 0$ the
 probability that
 \be \label{eq:probestoninfspec0}
  \big( I_n^+[B_{\omega_+}] -  (2n+1) \mu h , I_{n+1}^-[B_{\omega_-}] +
 (2n+3) \mu h  \big) \subset \rho(H(\wt{A}_\Lambda))
 \ee
 holds,
 satisfies the lower bound
\begin{align} \label{eq:probestoninfspec}
{\mathbb{P}} \left\{ \eqref{eq:probestoninfspec0} \right\}
  \geq 1 -  | \widetilde{\Lambda} | [ \nu_+(  c_u^{-1} h ) + \nu_-(  c_u^{-1} h ) ] .
\end{align}
\end{theorem}

For the sake of a transparent  exposition, we set
$( m_+)_z^{(k)} := ( \omega_+)_z^{(k)} $.

\noindent
\begin{proof}
Consider the events
\begin{align}
& \omega^{0}_z  \leq m_+^{(0)}  - c_u^{-1} h   , \quad z \in \wt{\Lambda}
\label{eq:zeroscalelif+} \\
&  \omega^{0}_z  \geq m_-^{(0)}  +  c_u^{-1} h  ,  \quad z \in \wt{\Lambda}  \label{eq:zeroscalelif-} .
\end{align}
If event \eqref{eq:zeroscalelif+} holds, then
\begin{eqnarray*}
   (2n+1) \wt{B}_\Lambda(x) + V(x)           & = &
 (2 n + 1 ) \Big[ B_{\rm det}(x)
 + \mu \sum_{k=0}^\infty \sum_{z \in \widetilde{\Lambda}^{(k)}} \omega_z^{(k)} u(x- z ) \Big] + V(x) \\
& \leq &  (2 n + 1 ) \Big[  B_{\rm det}(x)          - \mu  \sum_{z \in \widetilde{\Lambda}^{(0)}} c_u^{-1} h   u(x- z )
 + \mu \sum_{k=0}^\infty \sum_{z \in \widetilde{\Lambda}^{(k)}} m_+^{(k)}  u(x- z )  \Big] + V(x)\\
& \leq &   (2 n + 1 ) \Big[ B_{\rm det}(x)       - \mu  h    + \mu \sum_{k=0}^\infty
\sum_{z \in \widetilde{\Lambda}^{(k)}} m_+^{(k)}  u(x- z ) \Big] + V(x)  \\
&\leq &   e_{n,{\rm max}}(B_{\omega_+} )  - (2 n + 1)  \mu  h   ,
\end{eqnarray*}
where we used  \eqref{eq:profileidentity1} from the second to third line.
By  this and the definition in  \eqref{eq:intofres} it follows that
the event  \eqref{eq:zeroscalelif+} implies
\be \label{eq:specleft}
I_n^+[\wt{B}_\Lambda] \leq I_n^+[B_{\omega_+}] - (2 n + 1 ) \mu h .
\ee
Similarly  one can show that the event \eqref{eq:zeroscalelif-} implies
\be \label{eq:specright}
I_{n+1}^-[\wt{B}_\Lambda] \geq I_{n+1}^-[B_{\omega_-}] + (2 n + 3 ) \mu h .
\ee
On the other hand,     by   Theorem \ref{thm:boundonspec}  it follows    that
\eqref{eq:specleft} and \eqref{eq:specright} imply \eqref{eq:probestoninfspec0}.
Thus it remains to estimate the following probability,
\begin{align*}
{\mathbb{P}}\{\eqref{eq:zeroscalelif+} \ \  {\rm and } \  \ \eqref{eq:zeroscalelif-} \}&=
\left[
{\mathbb{P}} \{ \omega_0^{(0)} \leq  m_+^{(0)} -  c_u^{-1} h
\ \ {\rm and } \ \    \omega_0^{(0)} \geq  m_-^{(0)} +  c_u^{-1} h   \}
\right]^{ | \widetilde{\Lambda} | }  \nonumber  \\
& \geq 1 -  | \widetilde{\Lambda} | ( \nu_+( c_u^{-1} h ) + \nu_-( c_u^{-1} h )  ) , 
\end{align*}
where the last line follows from the binomial formula.
\end{proof}

\begin{corollary}\label{lem:probestoninfspec2}   Assume that
{\bf ($\boldsymbol{{\rm A}_\tau}$)} holds for some fixed
 $\tau > 2$ and   $c_v$.
 For any $\xi \in (0,  \tau - 2)$ set
 $ \beta := \frac{1}{2}(1- \frac{\xi+2}{\tau}) \in (0,1)$, then
there is  an $l_{\rm initial} = l_{\rm initial}(\tau,  \xi,c_u,c_v,c_\delta)$ such that
\begin{align*}
{\mathbb{P}} \left\{  {\rm dist}( \sigma (H(\wt{A}_\Lambda ) ,
 ( I_n^+[B_{\omega_+}]  , I_{n+1}^-[B_{\omega_-}] ) ) \geq  ( 2 n + 1 ) \mu l^{\beta-1}  \right\}
  \geq   1 -  \frac{1}{4}l^{-\xi}
\end{align*}
for any $\Lambda = \Lambda_l(x)$, with $x \in \Z^2$ and $l \geq l_{\rm initial}$ (we adopt the convention
that the distance to the empty set is infinity).
\end{corollary}

\medskip

\noindent
\begin{proof}
 Set
$h = l^{ \beta - 1 }$ in Theorem
\ref{lem:probestoninfspec}. Then
$$| \widetilde{\Lambda} | [\nu_+( c_u^{-1} h ) + \nu_-( c_u^{-1} h ) ]\leq  |\widetilde{\Lambda}| 2 c_v (c_u^{-1} h)^\tau
 =  2 c_u^{-\tau}  c_v (l+ c_\delta)^{ 2} l^{    (\beta - 1 )  \tau } \leq \frac{1}{4}l^{-\xi}, $$
where the first inequality follows from assumption {\bf ($\boldsymbol{{\rm A}_\tau}$)},
and the  second inequality holds  for large $l$.
\end{proof}

\subsection{Multi-scale Analysis:  Proof of Theorem  \ref{thm:loclization} } \label{sec:multiscale}

For the proof of Theorem  \ref{thm:loclization}, we will essentially
follow the argument in \cite{EH2} that is based
on \cite{S}  after including the magnetic field.

We assume  {\bf ($\boldsymbol{{\rm A}_\tau}$)} throughout this section  for some fixed
 $\tau > 2$ and   $c_v$.
The constants $b_0,  \rho, \delta$ are as in the assumptions of Theorem \ref{thm:wegner}.

\begin{definition}
A square $\Lambda$ is called {\bf $(\gamma,E)$-good} for $\omega \in \Omega$ if
$$
\| \chi^{\rm out}  (H_{\Lambda}(A_\omega) - E )^{-1}  \chi^{\rm int}  \| \leq \exp(-\gamma l ) ,
$$
where  $E \in \rho(H_\Lambda(A_\omega))$.
\end{definition}

Let us introduce the  multiscale induction hypotheses.
Below we denote by   $I \subset \R$   an interval and assume $l \in 2 \N + 1$.
First,
for  $\gamma > 0$, and $\xi > 0$ we introduce  the following  hypothesis.

\medskip

\noindent
$G(I,l,\gamma,\xi)$:
\quad
$\forall x, y \in \Z^2$, $| x - y |_\infty \geq l+c_\delta $, the following estimate holds:
$$
{\mathbb{P}} \{ \forall E \in I  , \ \Lambda_l(x) \ {\rm or } \  \Lambda_l(y)
\ {\rm is } \ (\gamma,E){\rm -good \ }  \} \geq  1 - l^{-2 \xi} .
$$

\smallskip\noindent
Note that this definition  includes a security
distance $c_\delta$, to ensure the independence
of  squares.

\begin{lemma} \label{lem:G} Fix $n \in \N$. For any  $\xi \in (0, \frac{\tau}{2} - 1)$ there is an
$l_G = l_G(\tau,\xi,c_u,c_v,c_\delta, K_0, K_1, \delta, \mu, \kappa, \rho,n)$ such that for all $ l \geq l_G b_0^\kappa$,  $G(I,l,\gamma,\xi)$
holds with $\gamma= l^{\beta-1}$, $\beta =    \frac{1}{2}(1- \frac{2 \xi+2}{\tau})        \in (0,1)$, and
if $I$ is any of the intervals of the form
\begin{align*}
I &=  I_n^+[B_{\omega_+}]  + [ - \frac{1}{2} \mu l^{\beta-1}, 0 ] \\
I &=  I_n^-[B_{\omega_-}]   + [0 ,  \frac{1}{2} \mu l^{\beta-1} ] ,
\end{align*}
as long as  $I \subset [ \frac{b_0}{2} ,  K_1 b_0]$.
\end{lemma}

\medskip

\noindent
\begin{proof}
First we give a  deterministic estimate and then we estimate the  probability.

\medskip

Let $A$ be a vector potential and let  $\Lambda = \Lambda_l$.
Suppose the vector potential satisfies
 \begin{align} \label{lem:eq:boundonbeta}
  {\rm dist}( \sigma ( H(\widetilde{A}_{\Lambda}) ) , I_{n}^{\pm}[B_{\omega_\pm} ] ) \geq  ( 2 n + 1 )  \mu l^{\beta-1} .
\end{align}
If $E \in I$, then ${\rm dist}( H(\wt{A}_\Lambda) , E) \geq ( 2 n + 1 ) \frac{1}{2} \mu l^{\beta-1} $.
Thus by the resolvent decay estimate, see Theorem  \ref{thm:combthomas}, we find
\be \label{Gdecay}
\| \chi^{\rm int} ( H_\Lambda(A)  - E)^{-1} \chi^{\rm out} \|
\leq  \frac{2 c_1}{\mu (2n+1)}   l^{1- \beta}  \exp(
 - c_2  ( \frac{1}{2}(2n+1) \mu l^{\beta-1}  )^{1/2}  l /4  ) ,
\ee
for $l \geq 4$. On the other hand if $\Lambda$ is $(E, b_0 C_\infty,\alpha)$-balanced, then
\be \label{Gdecay2}
\left\| \chi^{\rm out} ( H_\Lambda(A) - E)^{-1} \chi^{\rm in} \right\| \leq ( 1 + b_0 C_\infty l^{\alpha} ) \left| \, {\rm r.h.s. \ of } \ \eqref{Gdecay} \ \right| .
\ee
Thus we conclude from \eqref{Gdecay} and \eqref{Gdecay2} that there
 exists an $\hat{L}_G  = \hat{L}_G(\tau,\xi,\mu,n,\alpha,\kappa)$, such that
$\Lambda$ is $(\gamma,E)$--good provided
\eqref{lem:eq:boundonbeta} holds, $\Lambda$ is  $(E, b_0 C_\infty,\alpha)$-balanced,
and
\be \label{condonl}
l \geq \hat{L}_G   b_0^\kappa .
\ee

\medskip

It remains to estimate the probability.
Let $\alpha > 3 C_1 (\kappa^{-1} + \rho)$ (with $C_1$ as in Theorem  \ref{lem:probestoninfspec}) and
let $l$ satisfy  \eqref{condonl}. Then using the conclusion of the sentence leading up to \eqref{condonl},
we find
\begin{eqnarray}
\lefteqn{
\mathbb{P}\big\{\forall E \in I , \, \Lambda_l(x)  \ {\rm or} \ \Lambda_l(y)
 \ {\rm is } \ (\gamma,E)-{\rm good} \big\} }  \label{multprobest0}  \\
&& \geq \mathbb{P}\big\{ \forall E \in I , \, \Lambda_l(x)  \ {\rm or} \ \Lambda_l(y) \ {\rm is } \ (E,b_0 C_\infty, \alpha))-{\rm balanced}
\ \  {\rm and} \ \eqref{lem:eq:boundonbeta} \ {\rm holds \ for } \ \Lambda_l(x) \ {\rm and } \ \Lambda_l(y) \big\}  \nonumber
\\
&& \geq 1 -   \mathbb{P}\big\{ {\rm not }[ \forall E \in I , \, \Lambda_l(x)
  \ {\rm or} \ \Lambda_l(y) \ {\rm is } \ (E,b_0 C_\infty, \alpha))-{\rm balanced}]\big\} \nonumber \\
&& \ \
- \mathbb{P}\big\{ {\rm not} [  \eqref{lem:eq:boundonbeta} \ {\rm holds \ for } \ \Lambda_l(x) \ {\rm and  } \ \Lambda_l(y) ] \big\}  \nonumber \\
&& \geq    p_{l,x} p_{l,y} - p_1    \label{multprobest} ,
\end{eqnarray}
where we set
\be \label{eq:weginiproof1}
p_1 := \mathbb{P}\Big\{ {\rm not }[ \forall E \in I , \, \Lambda_l(x) 
 \ {\rm or} \ \Lambda_l(y) \ {\rm is } \ (E,b_0 C_\infty, \alpha))-{\rm balanced}]\Big\}
\ee
and we used that  by independence
$$
\mathbb{P}\Big\{  {\rm not} [  \eqref{lem:eq:boundonbeta} \ {\rm holds \ for } \ \Lambda_l(x) \ {\rm and  } \ \Lambda_l(y) ]  \Big\} = 1 - p_{l,x} p_{l,y}
$$
where
$$
p_{l,x} = p_{l,y} = p_{l,z} :=  \mathbb{P}\big\{
  \eqref{lem:eq:boundonbeta} \ {\rm holds \ for } \ \Lambda_l(z)   \big\} .
$$
It remains to estimate \eqref{multprobest}.
Observe that  by Corollary \ref{lem:probestoninfspec2} with ($\xi \to 2 \xi$) there exists an
$l_{\rm initial} = l_{\rm initial}(\tau,  \xi,c_u,c_v,c_\delta)$  such that for all
$l \geq l_{\rm initial}$
\be \label{ineq:lifasy}
p_{l,x} \geq 1 - \frac{1}{4} l^{- 2 \xi} .
\ee
On the other hand by Lemma  \ref{lem:wegini}
\be \label{eq:weginiproof2}
p_1 \leq  3 C_0^2   \mu^{-3}  l^{- \alpha/3}  \leq \frac{1}{4}l^{- 2\xi} ,
\ee
where we  used that the width of the interval  for which we want to prove
localization is bounded by $\mu$, and the last inequality follows if we choose
 $\alpha$ sufficiently large.
Thus inserting \eqref{eq:weginiproof2} and \eqref{ineq:lifasy} into  \eqref{multprobest},
we find that  \eqref{multprobest0} is bonded from below by
$$
1 - \frac{1}{4} l^{-2\xi}  - \frac{1}{2} l^{-2\xi}  
\geq 1 - l^{- 2\xi} ,
$$
for $l \geq l_{\rm initial}$ satisfying \eqref{condonl}.
\end{proof}

\medskip

For $\Theta > 0$, and $q > 0$ we introduce  the following hypothesis.
\medskip

\noindent
$W(I,l,\Theta,q)$:
\quad
For all $E \in I$ and $\Lambda = \Lambda_l(x)$, $x \in \Z^2$, the following estimate holds:
$$
{\mathbb{P}} \big\{  {\rm dist}(\sigma(H_\Lambda(A)), E)  \leq \exp(-l^\Theta) \big\} \leq l^{- q} .
$$

\medskip

The following Lemma is a consequence of  Theorem \ref{thm:wegner}.

\medskip

\begin{lemma} \label{lem:W} Suppose the assumptions of Theorem \ref{thm:wegner} hold.
Let  $\Theta > 0$,  $q > 0$, and $0 < \kappa \leq 1$. Let  $I \subset \R$ be a finite interval with
$\inf I \geq b_0/2$.
Then there exists a  constant
$l_W^* = l_W^*(I,\Theta,q,K_0,K_1,\delta,\mu,\kappa,\rho)$
such that
$W(I,l,\Theta,q)$ holds for all $l \geq l_W^* b_0^{\kappa}$.
\end{lemma}

\medskip

\noindent
{\bf Proof of Theorem \ref{thm:loclization}.} We consider only the upper band edges. The lower band edges are proven analogously.
 Fix  $\xi \in (0, \tau - 2)$ and let $\beta = \frac{1}{2}(1- \frac{\xi+2}{\tau})$.
Choose, $0 < \Theta < \beta/2$ and $q > 2$ and 
set $\kappa :=\frac{1}{2} \min( (2 - 2 \beta)^{-1},1)$.
By  Lemma  \ref{lem:G} there exists  an
${l}_{G} = l_G(\tau,\xi,c_u,c_v,c_\delta, K_0, K_1, \delta, \mu, \rho)$ 
such that
$G(I_{l},l,\gamma_{l},\xi)$ holds  with
$I_l =    I_n^+[B_{\omega_+}]   + [ - \frac{1}{2} \mu l^{\beta-1}, 0 ]$
and $\gamma_{l} := l^{\beta-1}$ for all $l \geq {l}_{G}b_0^\kappa$.
By Lemma \ref{lem:W}  there exists an
${l}_{W}^* $ (depending on  $\Theta,  q, K_0, K_1, \delta, \mu, \kappa , \rho$)
such that  $W( I_{l}, l , \Theta, q  )$ is satisfied for
$l \geq l_{W}^* b_0^{\kappa}$ and thus also for
$l \geq l_0 := \max(l_{W}^* b_0^{\kappa},l_{G}b_0^{\kappa})$.
Thus one now apply the multiscale analysis as outlined in
\cite{S} for the interval $J_0 := I_{l_0}$ (This is explained in detail in \cite{EH2}).

Having established the application of the multiscale analysis we can now show  using standard arguments that   $H(A_\omega)$ has
pure point spectrum in $J_0$ for almost all $\omega \in \Omega$.
We write
$J_0 =   I_n^+[B_{\omega_+}] + [ - e_0, 0 ]$, with
$e_0 := \frac{1}{2} \mu l_0^{\beta-1}$.
Observe that by  Theorem \ref{thm:boundonspec} there is no spectrum in the interval
 $(  I_n^+[B_{\omega_+}]  ,  I_{n+1}^-[B_{\omega_-}]   )$.
Thus to conclude that  an  interval of the form  \eqref{eq:intofloc} contains pure point
spectrum, it suffices to show that $E_n^+ \in J_0$ and an analogous statement for the
corresponding lower band edge. To this end note that we have
\be \label{somebound}
  C_n ( K_2^\pm b_0^{-1/2} +  K_3^{\pm} b_0^{-2} )
 \leq \frac{1}{2} e_0  = \frac{1}{4}\mu b_0^{\kappa(\beta-1)}\big[\max (l_W^*,l_G)\big]^{\beta-1},
\ee
if $b_0$ is sufficiently large, noting that $b_0^{\kappa(\beta-1)}\ge b_0^{-1/4}$
by the choice of $\kappa$.
Since $K_3^\pm\le [K_2^\pm]^2$, the threshold for $b_0$ grows with $[K_2^\pm]^4$
for large $K_2^\pm$.
By \eqref{somebound} and the definition \eqref{eq:intofresesup} 
 (recall  \eqref{eq:equivdef}) it
now  follows that $E_n^+ \in J_0$.
In view of  Theorem \ref{thm:estonspeca} an analogous argument now implies \eqref{eq:thereisspec}.

\qed

\bigskip

\appendix

\section{Analytic properties of the resolvent $R(z)$}\label{appres}

We cite the following Lemma found in \cite{korpus10}.

\begin{lemma} \label{lem:korpus}
For any $z \in \C \setminus \sigma(H)$, the integral kernel of the resolvent $R(z) = (H-z)^{-1}$ of the magnetic
Hamiltonian $H = (p_1 + \frac{B_0}{2} x_2)^2 + (p_2 - \frac{B_0}{2} x_1)^2$, with constant magnetic field $B_0$,
can be expressed in terms of the $\Gamma(\cdot)$-function and confluent hypergeometric function
$U(\cdot,\cdot;\cdot)$ as follows:
\be
R(z)(x,y) = \frac{1}{4 \pi } \Gamma(w) U(w,1;\zeta) \exp(-\frac{1}{2} \zeta - i \frac{B_0}{2}[x,y]) ,
\ee
with $[x,y] =  x_1 y_2 - x_2 y_1$,
$w = \frac{1}{2} - \frac{ z }{2 B_0}$, and $\zeta = \frac{B_0}{2}|x-y|^2$.
\end{lemma}

Fix  $n \in \N_0$. Suppose   $z \in \C$  is in the \eqref{eq:neighlandau} neighborhood
of the $n$-th Landau level.
In terms of $w$ this  is equivalent to
\be \label{eq:wwindow}
-n - \frac{1}{2} \leq {\rm Re} w \leq -n + \frac{1}{2} , \quad |{\rm Im} w | \leq 1/2 .
\ee
It is well known that the $\Gamma$ function is a meromorphic function with simple poles at $0,-1,...$. Thus
\begin{equation} \label{eq:boundongamma}
| \Gamma( \frac{1}{2} - \frac{ z }{2 B_0}  ) | \leq C_n \left[  \frac{B_0}{|z - B_0 (2 n + 1 )|} + 1  \right] ,
\end{equation}
for some constant $C$ (depending on $n$).
Furthermore, there  exists a  constant $C$
(depending on $n$)  such that
\begin{align} \label{eq:boundonU}
U(w,1;\zeta) &\leq  C ( 1 + \ln  \zeta  +  \zeta^{n+1}) , \\
U_3(w,1;\zeta) &\leq  C ( 1 +  \zeta^{-1} +  \zeta^{n+1}) ,  \label{eq:boundonU2}
\end{align}
for all $\zeta > 0$.
This estimate  can be seen as follows. We use the the integral representation
 13.2.5. in \cite{AS},
 \be \label{eq:intrepU}
 \Gamma(a) U(a,1;\zeta) = \int_0^\infty e^{- \zeta t} t^{a-1} ( 1 + t )^{-a} dt , \qquad {\rm Re} a > 0 , \ \ {\rm Re} \zeta > 0 .
 \ee
Note that \eqref{eq:intrepU} can  a priori  only be applied  if ${\rm Re} a > 0$, but
we need it for  $w$ satisfying \eqref{eq:wwindow}.
To circumvent this problem, we iterate the   recurrence relation 13.4.15. in \cite{AS}
$$
U(a , 1 ; \zeta) =   (1+2 a+\zeta) U(a +1,1 ; \zeta) -  (a+1)^2 U(a+2,1 ; \zeta)  ,
$$
$n+1$ times, which yields
\be \label{eq:iteratedU}
U(w  , 1 ; \zeta ) = P_n(a,\zeta) U(w + n+1,1;\zeta) + Q_n(w ,\zeta) U(w+n+2,1;\zeta) ,
\ee
for some polynomials  $P_{n}$ and $Q_{n}$  with degree at most $n+1$ in $\zeta$ and
$2(n+1)$ in $w$. Now \eqref{eq:boundonU} and \eqref{eq:boundonU2} can be shown
using  \eqref{eq:intrepU} to estimate the
right hand side of  \eqref{eq:iteratedU} (observe that the $\Gamma$ function has
no zeros in the set  $[\frac{1}{2}, \frac{5}{2} ] + i[-\frac{1}{2}, \frac{1}{2}]$ ).
The large $\zeta > 0$ behavior in \eqref{eq:boundonU} is now  trivial to see.
For  the small $\zeta$ behavior note that  \eqref{eq:intrepU} diverges logarithmically
in $\zeta$ as $\zeta \downarrow 0$, and that its derivative with respect to $\zeta$
 diverges like $\zeta^{-1}$.

\section{Combes-Thomas decay estimate}

Define the function  $\rho(x) = ( 1 + |x|^2)^{1/2}$. Let $\widetilde{H}$ be an operator of the form
$H_\Lambda(A)$. Define
\begin{equation}
\widetilde{H}(\alpha) := e^{ i \alpha \rho} \widetilde{H} e^{- i \alpha \rho}
= \widetilde{H} -  \alpha \nabla \rho \cdot ( - i \nabla - a) -   ( - i \nabla - a)  \cdot \alpha \nabla \rho
  + \alpha^2 | \nabla \rho |^2 .
\label{eq:defofdecay}
\end{equation}
Since $|\nabla \rho|$ and $|\Delta \rho|$ are
bounded and $( - i \nabla - a)$ is infinitesimally small with respect to $\widetilde{H}$,
we obtain that $\widetilde{H}(\alpha)$ is an analytic family of type A on $\C$.
Using this property one can show the following result, following the proof of Theorem 2.4.1. in \cite{S}.

\begin{theorem} \label{thm:combthomas}  Let $R > 0$. Then there exists a $c_1=c_1(R)$  and $c_2=c_2(R)$ such that the following holds. $(r,s ) \subset \rho(H) \cap (-R,R)$, $E \in (r,s)$ and $\eta = {\rm dist}(E, (r,s)^c) > 0$ imply
the estimate
$$
\| {\bf 1}_B (H(A) - E)^{-1} {\bf 1}_D \|
\leq c_1 \eta^{-1} \exp[-c_2 (s-r)^{1/2} \eta^{1/2} \delta ] ,
$$
where  $B$ and $D$ denote a set in $\R^2$ and  ${\rm dist}(B,D) =: \delta > 0$.
\end{theorem}

\end{document}